\documentclass[runningheads]{llncs}
\pdfoutput=1
\usepackage{mwe}
\usepackage{graphicx}
\graphicspath{{figures/}}
\usepackage{color}
\definecolor{highlight}{rgb}{1,1,0.6}
\definecolor{link}{rgb}{0.5,0.0,0.0}
\definecolor{cite}{rgb}{0.0,0.0,0.6}
\definecolor{url} {rgb}{0.3,0.0,0.3}
\definecolor{grey}{rgb}{0.3,0.3,0.3}

\usepackage{tabularx}
\usepackage{array}
\usepackage{booktabs} 
\usepackage{arydshln} 

\usepackage{soul} 
\usepackage{xspace}
\usepackage[shortcuts]{extdash}
\usepackage{relsize} 

\sethlcolor{highlight}

\newcommand{\demon}{{DEMON}}
\newcommand{\infomap}{{Infomap}}

\usepackage{amssymb}
\usepackage{bm}
\usepackage{mathtools}

\usepackage{rotating}
\newcolumntype{P}[1]{>{\centering\arraybackslash}p{#1}}
\newcolumntype{Y}{>{\centering\arraybackslash}X}
\usepackage{subcaption}

\begin{document}
    \title{A customisable pipeline for continuously harvesting socially-minded Twitter users}
    \titlerunning{Harvesting socially-minded Twitter users}
    %
    \author{Flavio Primo\inst{1} \and
    Paolo Missier\inst{1} \and
    Alexander Romanovsky\inst{1} \and
    Mickael Figueredo\inst{2} \and
    Nelio Cacho\inst{2}}

    \authorrunning{F. Primo et al.}
    %
    \institute{Newcastle University, School of Computing \\
    Science Central, Newcastle upon Tyne, UK \\
    \email{\{firstname.lastname\}@ncl.ac.uk}  \and
    Universidade Federal do Rio Grande do Norte \\
    Natal/RN - Brasil\\
    \email{neliocacho@dimap.ufrn.br}}
    \maketitle       

    \begin{abstract}
    	On social media platforms and Twitter in particular, specific classes of users such as \textit{influencers}  have been given satisfactory operational definitions in terms of  network and content metrics.
    	Others, for instance \textit{online activists}, are not less important but their characterisation still requires experimenting.
    	We make  the hypothesis that such interesting users can be found within temporally and spatially localised \textit{contexts}, i.e., small but topical fragments of the network containing interactions about social events or campaigns with a significant footprint on Twitter.
    	To explore this hypothesis, we have designed a continuous user profile discovery pipeline that produces an ever-growing dataset of user profiles by harvesting and analysing contexts from the Twitter stream.
    	The profiles dataset includes key network and content-based users metrics, enabling experimentation with user-defined score functions that characterise specific classes of online users.
        The paper describes the design and implementation of  the pipeline and its empirical evaluation on a case study consisting of healthcare-related campaigns in the UK, showing how it supports the operational definitions of online activism, by comparing three experimental ranking functions. The code is publicly available.
    	
    \end{abstract}

    \section{Introduction}

We present a customisable software framework for incrementally discovering and ranking individual profiles for classes of online users, through analysis of their social activity on Twitter.
Practical motivation for this work comes from our ongoing effort to support health officers in tropical countries, specifically in Brazil, in their fight against airborne virus epidemics like Dengue and Zika. Help from community activists is badly needed to supplement the scarce public resources deployed on the ground. Our past work has focused on identifying relevant content on Twitter that may point health authorities directly to mosquito breeding sites~\cite{Sousa2018}, as well as to users who have shown interest in those topics, i.e., by posting relevant content on Twitter~\cite{Missier2017}. 

The approach described in this paper generalises those past efforts, by attempting to discover users who demonstrate an inclination \textit{to become engaged in social issues, regardless of the specific topic}.
We refer to this class of users as \textit{activists}.
The rationale for this approach is that activists who manifest themselves online on a range of social initiatives, may be more sensitive to requests for help on specific issues than the average Twitter user.
In the paper we experiment with healthcare-related online campaigns in the UK. Application of the approach to our initial motivating case study in ongoing as part of a long-term collaboration, and is not specifically discussed in the paper.

To be clear, this work is not about providing a robust definition of online activism, or to demonstrate that online activism translates into actual engagement in the ``real world''.
Instead, we start by acknowledging that the notion of activist is not as well formalised in the literature as that of, for example, \textit{influencers}, and we develop a generic content processing pipeline which can be customised to identify a variety of classes of users. 
The pipeline repeatedly searches for and ranks Twitter user profiles by collecting quantitative network- and content-based user metrics. 
Once targeted to a specific topic, it provides a tool for exploring operational definitions of user roles, including online activism, i.e., by combining the metrics into higher level, \textit{engineered} user features to be used for ranking.

Although the user harvesting pipeline is generally applicable to the analysis of a variety of user profiles, our focus is on the search for a satisfactory operational definition of online activism. 
According to the Cambridge Dictionary, an \textit{activist} is ``A person who believes strongly in political or social change and takes part in activities such as public protests to try to make this happen''.
While activism is well-documented, e.g. in the social movement literature~\cite{doi:10.1080/14742830701497277}, and online activism is a well-known phenomenon \cite{IJoC1246}, research has been limited to the study of its broad societal impact. 
In contrast, we are interested in the fine-grained discovery of activists at the level of the single individual, that is, we seek people who feel passionate about a cause or topic, and who take action for it. 
Searching for online activists is a realistic goal, as activists presence in social media is widely acknowledged, and it is also clear that social media facilitates activists communication and organization \cite{Poell2014,Youmans2012}. 
Specific traits that characterise activists include awareness of causes and social topic and the organization of social gatherings and activities, including in emergency situations, by helping organize support efforts and diffusion of useful information.

\subsection{Challenges}
 
The definition of online activism translates into technical challenges in systematically harvesting suitable candidate users.
Firstly, the potentially more subdue nature of activists, relative to influencers, makes it particularly difficult to separate their online footprint from the background noise of general conversations.
Also, interesting activists are by their nature associated to specific topics and manifest their nature in local contexts, for instance as organisers or participants to local events. 
Finally, we expect their personal engagement to be sustained over time and across multiple such contexts. 
These observations suggest that the models and algorithms developed for influencers are not immediately applicable, because they mostly operate on global networks, where less prominent users have less of a chance to emerge.
Some topic-sensitive metrics and models have been proposed to measure social influence, for example, \textit{alpha centrality}~\cite{Bonacich2001,Overbey2013} and the \textit{Information Diffusion} model~\cite{Pal2011}. Algorithms based on topic models have also been proposed to account for topic specificity~\cite{Zhao2011b}. However, these approaches are still aimed at measuring influence, not activism, and assume a one-shot discovery process, as opposed to a continuous, incremental approach.

\subsection{Approach and Contributions}

To address these challenges, the approach we propose involves a two-fold strategy. 
Firstly, we identify suitable contexts that are topic-specific and limited both in time and, optionally, also in space, i.e., regional initiatives, events, or campaigns.
We then search for users only within these contexts, following the intuition that low-key users who produce weak online signal have a better chance to be discovered when the search is localised and then repeated across multiple such contexts.
By continuously discovering new contexts, we hope to incrementally build up a users' dataset where users who appear in multiple contexts are progressively more strongly characterised.
Secondly, to allow experimenting with varying technical definitions of \textit{activist}, we collect a number of network-based and content-based user profile features, mostly known from the literature, and make them available to experiment with a variety of user rankings.

The paper makes the following specific contributions.
Firstly, we propose a data processing pipeline for harvesting Twitter content and user profiles, based on multiple limited contexts. 
The pipeline includes community detection and network analysis algorithms aimed at discovering users within such limited contexts.

Secondly, we have implemented a comprehensive set of content-based metrics that results into an ever-growing database of user profile features, which can then be used for mining purposes. 
User profiles are updated when they are repeatedly found in multiple contexts.

Lastly, for empirical evaluation of our implementation, we demonstrate an operational definition of the activist profile, defined in terms of the features available in the database. We collected about 3,500 users  across 25 contexts in the domain of healthcare awareness campaigns in the UK during 2018, and demonstrated three separate ranking functions, showing that it is possible to identify individuals as opposed to well-known organisations.
The application of the approach to the specific challenge of combating tropical disease epidemics in Brazil is currently in progress and is not reported in this paper.

\subsection{Related Work}  \label{sec:related}

The closest body of research to this work is concerned with techniques for the discovery of online \textit{influencers}. 
According to~\cite{Kardara2015}, influencers are \textit{prominent individuals with special characteristics that enable them to	affect a disproportionately large number of their peers with their actions}.
A large number of metrics and techniques have been proposed to make this generic definition operational~\cite{RIQUELME2016949}. 
These metrics tend to favour high visibility users across global networks, regardless of their actual impact~\cite{Cha2010MeasuringUI}. 
In contrast, activists are typically low-key, less prominent users who only emerge from the crowd by signalling high levels of engagement with one or more specific topics, as opposed to being thought-leaders. 
While such behaviour can be described  using well-tested metrics~\cite{RIQUELME2016949}, it should also be clear that new ways to combine those metrics are required.
A method for creating Twitter user ontologies considering the content type of the tweets is proposed in~\cite{6978961}. This approach could be used to gain insights over a user, but fails to give a comprehensive description of the user activity as it is based only on recent user activity, also due to Twitter API limitations.

The algorithm proposed in \cite{MATIC2011} aims to identify influencers based on a single topic context, based on relevant social media conversations.
Metrics include number of ``likes'', viewers per months, post frequency and  number of comments per post, as well as the ratio of positive to negative posts.
As some of these metrics are qualitative and difficult to acquire, however, this approach is not easy to automate.
Another approach to ranking topic-specific influencers within specific events appears in  \cite{Kardara2015}, where network dynamics are accounted for in real-time.
Once again however, the effect is to discover users who receive much attention, but do not necessarily create a real impact over users inside one topic.

Machine learning is used in~\cite{Biran2012} to analyse posted content and recognise when a user is able to influence another  inside a conversation.
This however requires substantial a priori ground truth, making this approach impracticable in our case. In addition, the need to create a classifier for each topic limits the scalability of the system.

A supervised regression approach is used in~\cite{7569535} to rank influence of Twitter users. It uses features that are not based on content, but the method performs poorly as it requires a huge training  set to work effectively.

	
Unlike the majority of the influencer ranking algorithms, in \cite{Schenk2011} a topic-specific influencer ranking is proposed. First it harvests sequentially timed snapshots of the network of users related to a topic. Then it ranks the users based on the number of followers gained and lost in the considered snapshots.

Finally, \cite{Bizid:2015:PUD:2808797.2809411} presents a model for identifying ``prominent users'' regarding a specific topic event in Twitter. Those are users who focus their attention and communication on the aforementioned topic event. Users are described by a feature vector, computed  in real-time, which allows a separation between on-topic and off-topic users activity over Twitter. Similar to~\cite{Biran2012}, problems of scalability and adaptability arise as two supervised learning methods are used, one to discriminate prominent users from the rest and the other to rank them.
    \vspace{-10pt}
\section{Contexts and user metrics}
\vspace{-5pt}


The aim of the pipeline is to repeatedly and efficiently discover user profiles from the Twitter post history within user-specified contexts and to use the process to grow a database of feature-rich user profiles that can be ranked according to user-defined relevance functions. 
The criteria used to define contexts, profile relevance functions, and associated user relevance thresholds can be configured for specific applications.

\subsection{Contexts and Context networks} \label{sec:contexts}
\vspace{-5pt}

A context $C$ is a Twitter query defined by a set $K$ of hashtags and/or keyword terms, a time interval $[t_1, t_2]$, and a geographical constraint  $s$, such as  a bounding box:
\begin{equation}
C = (K, [t_1, t_2], s)
\label{eq:context}
\end{equation}
Let $P(C)$ denote the query result, i.e., a set of posts by users.
We only consider two Twitter user activities: an \textit{original tweet}, or a \textit{retweet}.
Let $u(p)$ be the user who originated a tweet $p \in P(C)$.
We say that both $p$ and $u(p)$ are \textit{within context} $C$.
We also define the complement $\Tilde{P}(C)$ of $P(C)$ as the set of posts found using the same spatio-temporal constraints, but which do not contain any of the terms in $K$. More precisely, given a context $C'= ( s, [t_1, t_2], \emptyset )$ with no terms constraints, we define $\Tilde{P}(C) = P(C') \setminus P(C)$. 
We refer to these posts, and their respective users, as ``out of context $C$''.

$P(C)$ induces a user-user social network graph $G_C = (V,E)$ where $V$ is the set of all users who have authored any $p \in P(C)$: 
$V = \{ u(p) | p \in P(C) \}$, and a weighted directed edge $e = \langle u_1, u_2, w \rangle$ is added to $E$ for each pair of posts $p_1, p_2$ such that $u(p_1) = u_1, u(p_2) = u_2$ and 
either (i) $p_2$ is a retweet of $p_1$, or (ii) $p_1$ contains a mention of $u_2$.
For any such edge, $w$ is a count of such pairs of posts occurring in $P(C)$ for the same pair of users.

\subsection{User relevance metrics}  \label{sec:metrics}

We support metrics that are generally accepted by the community as forming a core, from which many different social user roles are derived~\cite{RIQUELME2016949}. 
We distinguish amongst three types of features, which differ in the way they are computed from the raw Twitter feed:
\begin{description}
	\item[Content-based metrics] that rely solely on content and \textit{not} on the user-user network. These metrics are defined relative to a topic of interest, i.e., a context;
	\item[Context-independent topological metrics] that encode context-independent, long-lived relationships amongst users, i.e., follower/followee; and 
	\item[Context-specific topological metrics] that encode user relationships that occur specifically within a context.
\end{description}

All metrics are functions of a few core features that can be directly extracted from Twitter posts. 
Given a context $C$ containing user $u$, we define:
\begin{align*}
\mathit{R1}(u) &\text{: Number of retweets by $u$, of tweets from other users in C;}\\
\mathit{R2}(u)&\text{: Number of unique users in $C$, who have been retweeted by $u$;}\\
\mathit{R3}(u)&\text{: Number of retweets of $u$'s tweets;}\\
\mathit{R4}(u)&\text{: Number of unique users in $C$ who retweeted $u$'s tweets;}\\
\mathit{P1}(u)&\text{: Number of original posts by $u$ within $C$;}\\
\mathit{P2}(u)&\text{: Number of web links found in original posts by $u$ within $C$;} \\
\mathit{F1}(u)& \text{: Number of followers of $u$;}\\
\mathit{F2}(u)& \text{: Number of followees of $u$}
\end{align*}
Note that, given $C$, we can evaluate some of the features above with respect to either $P(C)$ or  $\Tilde{P}(C)$ independently from each other, that is, we can consider an ``on-context'' and an ``off-context'' version of each feature, with the exception of $\mathit{F1}$ and $\mathit{F2}$ which are context-independent.
For example, we are going to write $R1_{on}(u)$ to denote the number of context retweets and $R1_{\mathit{off}}(u)$ the number of out-of-context retweets by $u$, i.e., these are retweets that occur within $C$'s spatio-temporal boundaries, but do not contain any of the hashtags or keywords that define $C$.  
We similarly qualify all other features.
Using these core features, the framework currently supports the following metrics.

\textbf{Content-based metrics}:
\begin{align}
\textit{Topical Focus:~\cite{Missier2017}:} ~ \mathit{TF}(u) & =  \frac{\mathit{P1}_{\mathit{on}}(u)}{\mathit{P1}_{\mathit{off}}(u) +1}    \label{eq:TF}\\
\textit{Topical Strength~\cite{Bizid2018}:} ~\mathit{TS}(u) & =	\frac{\mathit{P2}_{\mathit{on}}(u) \cdot \log(\mathit{P2}_{\mathit{on}}(u) + R3_{\mathit{on}} +1 )}{\mathit{P2}_{\mathit{off}}(u) \cdot \log(\mathit{P2}_{\mathit{off}}(u) + R3_{\mathit{off}} +1 ) + 1}   \label{eq:TS} \\
\textit{Topical Attachment~\cite{Bizid:2015,Poell2014}:} ~\mathit{TA}(u) & = \frac{\mathit{P1}_{\mathit{on}}(u) + \mathit{P2}_{\mathit{on}}(u)}{\mathit{P1}_{\mathit{off}}(u) + \mathit{P2}_{\mathit{off}}(u) +1} \label{eq:TA}
\end{align}

The framework supports one \textbf{Context-independent topological metric} and one \textbf{Context-specific topological metric}, both commonly used, see e.g.~\cite{RIQUELME2016949}:
\begin{align}
\textit{Follower Rank:}  \quad \mathit{FR}(u) = \frac{\mathit{F1}(u)}{\mathit{F1}(u)+\mathit{F2}(u)}   \label{eq:FR}\\
\textit{In-degree centrality:} \quad \mathit{IC}(u) = \frac{\mathit{indegree}(u)}{N-1}  \label{eq:IDC}
\end{align}
where $N$ is the number of nodes in the network induced by $C$.
Note that the metrics we have selected are a superset of those indicated in recent studies on online activism, namely \cite{IJoC1246} and \cite{Poell2014}, and thus support our empirical evaluation, described in Sec.~\ref{sec:evaluation}.

    \section{Incremental User Discovery} \label{sec:Pipeline}
	
\begin{figure*}
	\centering
	\includegraphics[width=\textwidth]{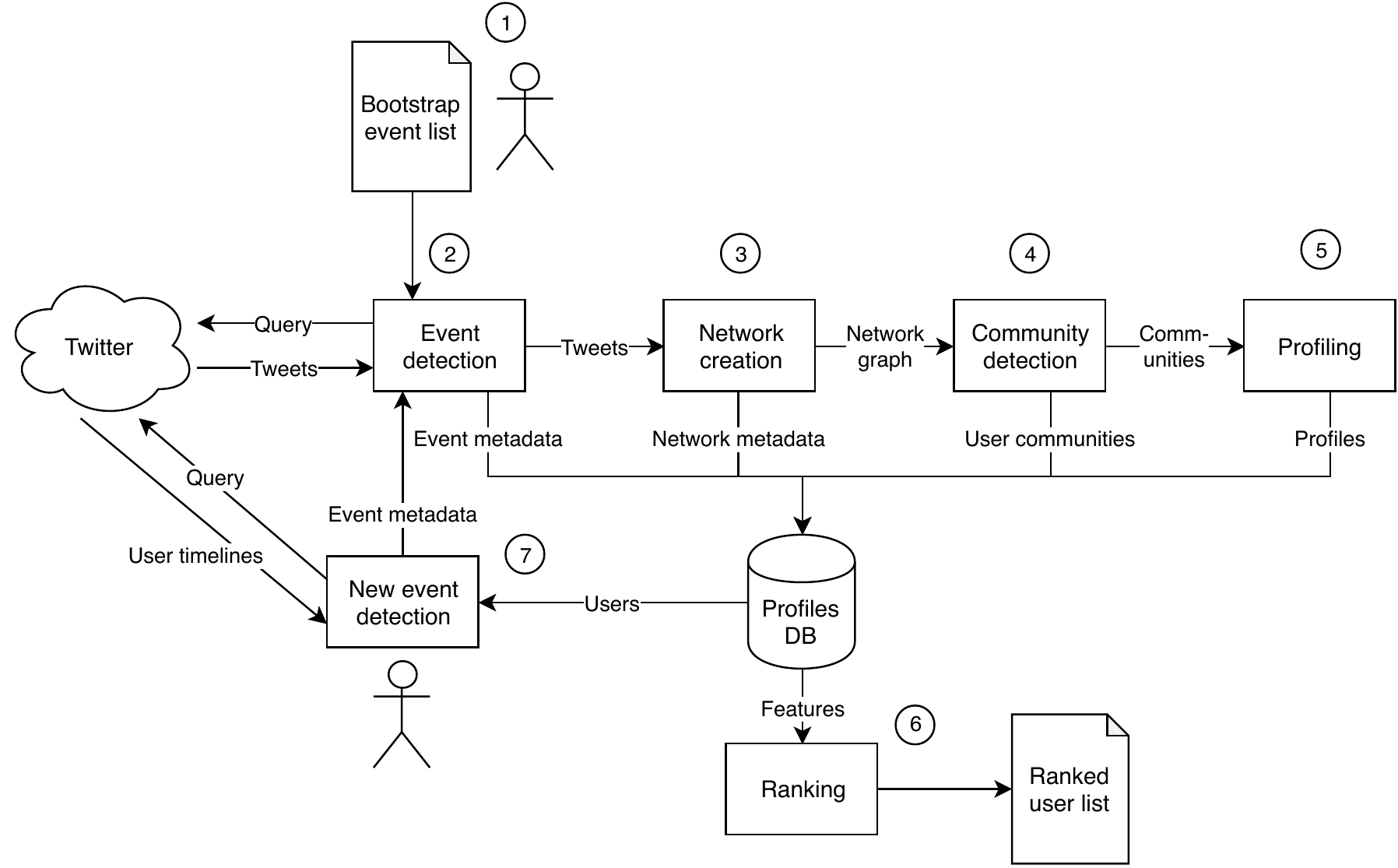}
	\caption{Schematic diagram of the user discovery framework. Note that an initial list $C$ of contexts (events) is provided to initialise the \textit{event detection} step. 
	The outputs from each of these steps are persisted into the Profiles DB.}
	\label{fig:twitterframework}
\end{figure*}

The content processing pipeline operates iteratively on a set of contexts within a given area of interest, for instance \textit{2018 UK health campaigns}. This set is initialised at the start of the process and then updated at the end of each iteration, in a semi-automated way. 
The user discovery process is therefore potentially open-ended, as long as new contexts can be discovered.
The new contexts are expected to be within the same topic area, but contexts that ``drift'' to new areas of interest are also acceptable. 
Each iteration takes a context $C$  as input, and selects a subset of the users who participate in $C$, using the topogical criteria described below, along with the  set of their features and metrics. 
These users profiles are added to a database, where entries for repeat users are updated according to a user-defined function. 
The pipeline structure is described below, where the numbers are with reference to Fig.~\ref{fig:twitterframework}.

%
Given $C$ as in (2), all Twitter posts $P(C)$ that satisfy $C$ are retrieved, using the Twitter Search APIs.
Note that this step hits the API service limitations imposed by Twitter. 
For this reason, in our evaluation we have limited our retrieval to 200 tweets/context. This is sufficient, considering that repeated users appear consistently in our evaluation (Sec.~\ref{sec:evaluation}). 
Twitter API limitations can be overcome by either extending the harvesting time, or by choosing more recent contexts, as the Twitter API is more tolerant with recents tweets.

The context network $G_C$ is then generated (3), as defined in Sec.~\ref{sec:contexts}.  
The size of each network is largely determined by the nature of the context, and  ranges between 140 and 400 users (avg 254, see Table~\ref{tab:contexts}).


Next, $G_C$  is partitioned into communities of users (4).
The goal of this partitioning is to further narrow the scope when computing in-degree centrality (\ref{eq:IDC}), 
to enable weak-signal users to emerge relative to other more globally dominant users.
We have experimented with two of the many algorithms for discovering virtual communities in social networks, namely \demon~\cite{Coscia:2012:DLD:2339530.2339630} and  \infomap~\cite{INFOMAP}. 
Both are available in our implementation, but based on our experimental comparison (Sec.~\ref{sec:evaluation}) we recommend the latter.

\demon~is based on \textit{ego networks}~\cite{Arnaboldi2013}, and uses a label propagation algorithm to assign nodes to communities.  
Users may be assigned to multiple communities, an attractive feature when users are active in more than one community within the same context, i.e., a social event or a campaign.
Label propagation is also a local method, translating into an efficient algorithm.
In practice, however, in our experiments we found that for almost half of our context networks, \demon~actually fails to discover any communities.
In contrast, \infomap~forces each user into at most one community, but it generates valid communities in all cases. 
As some of those are very small, our  implementation discards communities with fewer than 4 users (see Sec.~\ref{sec:evaluation}).
%

%



Once communities are identified, using either method, we calculate  in-degree centrality (\ref{eq:IDC}) for each node locally, either relative to their own community if they are available, and to the entire network otherwise.

\subsection{Computing user features and ranking} \label{sec:features}

Next, user metrics as defined in Sec.~\ref{sec:metrics}, along with the \textit{Follower Rank} are computed from the network and the user features.
This is achieved through bulk retrieval of user profile information (5), namely the number of tweets, retweets,  number of followers $F1(u)$ and followees, $F2(u)$, along with user name, web link, and bio.
Computing the other metrics: \textit{Topical Focus} (\ref{eq:TF}), \textit{Topical Strength} (\ref{eq:TS}), \textit{Topical Attachment} (\ref{eq:TA}) also requires the entire user post history to be retrieved for the entire time interval defined by the context.
These posts are then separated into $P(C)$ (on-context) and $\Tilde{P}(C)$ (off-context), depending on whether they contain a hashtag related to the context or not.
Similarly, a post that contains a link is a \textit{link on-topic} if it contains both a link and a hashtag related to the context, and a \textit{link off-topic} otherwise.
We also calculate the number of retweets for every post, i.e., $\mathit{R1}(u)$ and $\mathit{R3}(u)$, which are required to compute \textit{Topical Strength}.

All of these features are persisted to a database which is made available for ranking purposes.
User-defined functions can be specified to update the Rank of pre-existing users, e.g. by combining scores assigned at different times.
The DB enables user-defined scoring functions, which result in user ranking lists (6). Examples of these are given later in Sec.~\ref{sec:evaluation}.
This framework approach is consistent with the experimental nature of our search for \textit{activists}, which requires exploring a variety of ranking functions.

\subsection{New contexts discovery} \label{sec:context-discovery}

The final step within one iteration (7) aims to discover new contexts, so that the process can start again (2).
Intuitively, once a score function has been applied and users have been ranked, we can hope to discover new interesting keywords and hashtags by exploring the timeline of the top-$k$ users.
Specifically,  we consider each hashtag found in the timelines, which is related to the broader topic and not yet considered in past iterations.
Each stored hashtag is then enriched with the information needed to perform a new iteration of the pipeline, namely (i) the temporal and spatial information of the context, and (ii) related hashtags.
Currently this step is only semi-automated, as making a judgement on the relevance of the new terms requires  human expertise.
While automating this step is not straightforward, this is not a very time-consuming step, and one can imagine an approach where such task is crowdsourced.

While the process ends naturally when no new contexts are uncovered from the previous ones, the system continuously monitors the Twitter stream for recent contexts. These may typically include events that are temporally recurring, and use similar hashtags for each new edition. In this case, their relevance is assessed on the basis of their past history.
    \section{Empirical Evaluation} \label{sec:evaluation}

Existing methods to discover specific classes on online users are typically validated using a supervised approach, i.e., they rely on expert-generated ground truth.
Such approaches, however, are vulnerable to the subjectivity of the experts, whereby the evaluation would be measuring the fit of the model to the specific experts' own assessment of user instances' relevance. 
In contrast, we follow an unsupervised approach with no a priori knowledge of user relevance. We aim to demonstrate the value of our pipeline in creating a database of online profiles that are ready to be mined, along with examples of candidate user ranking functions.
In this approach, human expertise only comes into play to assess and validate the top-$k$ user lists produced by these functions.
We  demonstrate the pipeline in action on a significant set of 25 initial contexts, and define three alternative ranking functions aimed at capturing the empirical notion of  \textit{online activists}.
The pipeline is fully implemented in Python using Pandas and public libraries (NetworkX, Selenium) and is available on github\footnote{ \url{https://github.com/flaprimo/twitter-network-analysis}}. 
All experiments are performed on a single Azure node with standard commodity configuration.
Note that we do not focus on system performance as all components operate in near-real time. One exception is  Twitter content harvesting, which is limited by the Twitter API and requires approximately 2 hours per context.

\subsection{Contexts and networks} \label{sec:contexts-selection}
 
We have manually selected 25 contexts within the scope of health awareness campaigns in the UK, all occurring in 2018 and well-characterised using predefined hashtags.
Due to limitations imposed by Twitter on the number of posts that can be retrieved within a time interval, only $200$ tweets were retrieved from each context.
 Table~\ref{tab:contexts} lists the events along with key metrics for their corresponding user-user networks. 
To recall, \textit{assortativity} measures how frequently nodes are likely to connect to other nodes with the same degree ($>0$) or with a different degree ($<0$). 
Negative figures (mean: -0.22, std dev: 0.17) are in line with what is observed on the broader Twitter network~\cite{Fisher2017}.
The very small figures for density, defined as $\frac{\#edges }{\mathit{\mathit{\#nodes}} \cdot (\mathit{\#nodes} -1)}$ (mean: 0.004, std dev: 0.002), suggest very few connections exist amongst users within a context. 
This makes it difficult to detect meaningful communities, as described below, thus for some contexts the topological metrics are measured on the entire network as opposed to within each community.
This view is also supported by the average node degree (mean: 2.04, std dev: 0.46) and the ratio of strongly connected components to the number of nodes (mean: 0.98, std. dev. 0.02).

\begin{table}
	\tiny
	\resizebox{\textwidth}{!}{
	    \begin{tabularx}{\textwidth}{|X|P{2.2cm}|P{1.2cm}|P{1cm}|P{1.4cm}|P{1.2cm}|P{1.3cm}|}

\hline
\textbf{Context name} & \textbf{Period (2018)} & \textbf{Nodes} & \textbf{Edges} & \textbf{Density} & \textbf{Avg degree} & \textbf{Assor-tativity} \\ \hline
16 days of action & 11-25 / 12-10 & 396 & 349 & 0.002 & 1.8 & -0.1 \\ \hline
Elf day & 12-03 / 12-12 & 365 & 436 & 0.003 & 2.4 & -0.2 \\ \hline
Dry january & 01-01 / 01-31 & 235 & 234 & 0.004 & 2.0 & -0.3 \\ \hline
Cervical cancer prevention week & 01-21 / 01-27 & 209 & 192 & 0.004 & 1.8 & -0.1 \\ \hline
Time to talk day & 02-06 / 02-07 & 268 & 231 & 0.003 & 1.7 & -0.2 \\ \hline
Eating disorder awareness week & 02-25 / 03-03 & 256 & 241 & 0.004 & 1.9 & -0.2 \\ \hline
Rare disease day & 02-28 / 03-01 & 294 & 206 & 0.002 & 1.4 & -0.2 \\ \hline
Ovarian cancer awareness month & 03-01 / 03-31 & 215 & 202 & 0.004 & 1.9 & -0.4 \\ \hline
Nutrition and hydration week & 03-11 / 03-17 & 273 & 326 & 0.004 & 2.4 & -0.3 \\ \hline
Brain awareness week & 03-11 / 03-17 & 307 & 281 & 0.003 & 1.8 & -0.1 \\ \hline
No smoking day & 03-13 / 03-14 & 254 & 219 & 0.003 & 1.7 & -0.3 \\ \hline
Epilepsy awareness purple day & 03-26 / 03-27 & 306 & 252 & 0.003 & 1.6 & -0.2 \\ \hline
Experience of care week & 04-23 / 04-27 & 176 & 196 & 0.006 & 2.2 & -0.1 \\ \hline
Brain injury week & 05-01 / 05-31 & 238 & 306 & 0.005 & 2.6 & -0.1 \\ \hline
Mental health awareness week & 05-14 / 05-20 & 268 & 245 & 0.003 & 1.8 & -0.5 \\ \hline
Dementia action week & 05-21 / 05-31 & 300 & 300 & 0.003 & 2.0 & -0.0 \\ \hline
Mnd awareness month & 06-01 / 06-30 & 141 & 234 & 0.012 & 3.3 & -0.3 \\ \hline
Wear purple for jia & 06-01 / 06-30 & 165 & 245 & 0.009 & 3.0 & -0.5 \\ \hline
Carers week & 06-11 / 06-17 & 270 & 277 & 0.004 & 2.1 & 0.0 \\ \hline
National dementia carers & 09-09 / 09-10 & 184 & 177 & 0.005 & 1.9 & -0.2 \\ \hline
Mens health week & 06-11 / 06-17 & 264 & 214 & 0.003 & 1.6 & -0.2 \\ \hline
Stress awareness day & 11-07 / 11-08 & 293 & 209 & 0.002 & 1.4 & -0.2 \\ \hline
National dyslexia week & 10-01 / 10-07 & 229 & 235 & 0.004 & 2.1 & -0.2 \\ \hline
Ocd awareness week & 10-07 / 10-13 & 202 & 193 & 0.005 & 1.9 & -0.6 \\ \hline
Jeans for genes day & 09-21 / 09-22 & 246 & 325 & 0.005 & 2.6 & -0.2 \\ \hline

\end{tabularx}

	}
	\caption{List of contexts used in the experiments along with network metrics.}
	\label{tab:contexts}
\vspace{-20pt}
\end{table}

\subsection{Communities}  \label{sec:communities-eval}

 \demon~and \infomap~ produce significantly different communities in each network. 
\demon~identifies communities in only 48\% of the networks, with an average of only 1.92 communities per network and 
a slightly negative  (-0.28) average assortativity  per community, in line with the average for their parent networks.
Only the users who belong to one of those communities, about 6\%, are added to the database.
For the remaining 52\% of networks where no communities are detected, users' in-degrees are calculated using the entire network, and all users are added to the database, 
for a total of 3,570 users being added to the database in our experiments using \demon.

In contrast, \infomap~provides meaningful communities for all networks.
Those with fewer than 3 users are discarded, leaving  18.88 communities per network on average, with 8.5 users per community on average.
When using Infomap, 3,567 users were added to the database (on average 253 users per network).
The average assortativity across all communities is again slightly negative (-0.43).
Table~\ref{tab:demon-vs-infomap} compares the two approaches on the key metrics just discussed. On the basis of this comparison, we recommend using \infomap, which we have used for  our evaluation.

\begin{table}[t]
		\footnotesize
	\resizebox{\textwidth}{!}{
	    \begin{tabularx}{\textwidth}{|X|P{1.7cm}|P{1.7cm}|}
\hline
\textbf{Metric} & \textbf{\demon} & \textbf{\infomap} \\ \hline
Fraction of networks with null communities & 0.52 & 0.0 \\ \hline
Number of communities per context (avg) & 1.92 & 18.88 \\ \hline
Fraction of network users added to the DB  (avg) & 0.06 & 0.59 \\ \hline
Fraction of repeat users  added to the DB across networks & 0.28 & 0.37 \\ \hline
\end{tabularx}
	}
	\caption{Comparing \demon~to \infomap~for community detection.}
	\label{tab:demon-vs-infomap}
\end{table}

	\vspace{-10pt}
\subsection{Users discovery}  \label{sec:users}
	\vspace{-10pt}
	
\begin{figure*}[htb]
	\centering
	\includegraphics[width=1.2\textwidth]{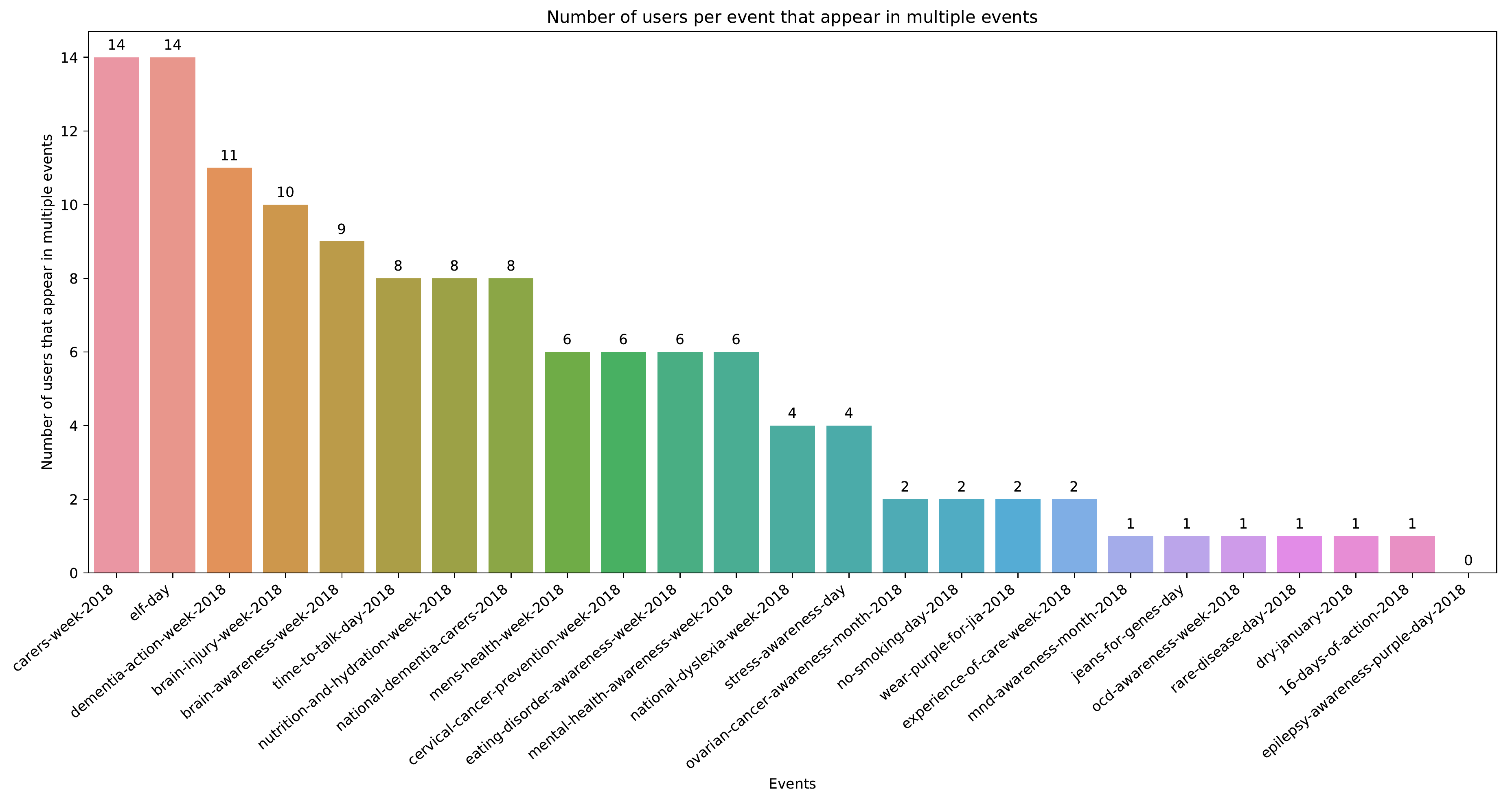}
	\caption{Number of repeat users for each context}
	\label{fig:repeat-users-frequency}
\end{figure*}

Repeat users who appear in multiple contexts are particularly interesting as they provide a stronger signal. 
Out of the total 3,567 users, 160 of those appear at least in two of the 25 contexts.
After community detection, only 61 of these users are still seen as repeat users,
while the remaining 99 are either removed altogether, or they only appear once.
Of the 61, 57 appear twice, 2 appear three times, and 2 appear four times. 
Thus, only 1.6\% of users appear more than once when communities with more than 3 users are considered, compared to the overall 4.5\% of overall repeat users.
Table~\ref{tab:repeat-users} reports the top-10 repeat users along with their \textit{Follower Rank}, and Fig.~\ref{fig:repeat-users-frequency} shows the number of repeat users per context.
As the table is sorted by number of occurrences then by \textit{Follower Rank}, an indication of popularity,  it is not surprising to find that top users include well-known names such as Mr. Hunt, who at the time of the events was Secretary of State for Health and Social Care in the UK, with $FR =1$, and a number of associations and foundations active in the public healthcare space.
More interesting are perhaps non-repeat users who emerge when ad hoc ranking is applied to the database, as we illustrate next.
\vspace{-10pt}
\begin{table}[htb]
	\centering
	\footnotesize
	\resizebox{\textwidth}{!}{
		\begin{tabularx}{\textwidth}{|p{2.5cm}|X|P{2.4cm}|P{2.4cm}|}
\hline
\textbf{Username} & \textbf{Name} & \textbf{Follower rank} & \textbf{Participations} \\ \hline
alzheimerssoc & Alzheimer's Society & 0.99 & 4 \\ \hline
dementiauk & Dementia UK & 0.98 & 4 \\ \hline
mentalhealth & Mental Health Fdn & 0.97 & 3 \\ \hline
colesmillerllp & Coles Miller LLP & 0.65 & 3 \\ \hline
jeremy\_hunt & Jeremy Hunt & 1.0 & 2 \\ \hline
nhsengland & NHS England & 0.99 & 2 \\ \hline
carersuk & Carers UK & 0.95 & 2 \\ \hline
rdash\_nhs & RDaSH NHS FT & 0.88 & 2 \\ \hline
alzsocseengland & Alzheimer's Society - South ... & 0.64 & 2 \\ \hline
mndassoc & MND Association & 0.64 & 2 \\ \hline
\end{tabularx}
	}
	\caption{Top-10 repeat users, amongst those who belong to a community.}
	\label{tab:repeat-users}
	\vspace{-20pt}
\end{table}

\vspace{-15pt}
\subsection{Users ranking} \label{sec:ranking}
\vspace{-5pt}

To demonstrate the potential value of the database, albeit on a small scale, we have tested three user ranking functions.
As mentioned, the aim of this exercise is to provide an objective grounding for engaging with experts on finding suitable operational definitions for specific user profiles. We consider good functions those that privilege individuals over organisations or business.
\begin{table}
	\centering
	\footnotesize
	\resizebox{\textwidth}{!}{
		\begin{tabular}{P{0.6cm}|p{2.3cm}|P{1.4cm}|P{1.8cm}|p{2.3cm}|P{1.4cm}|P{1.8cm}|p{2.3cm}|P{1.4cm}|P{1.8cm}|}

\cline{2-10}
\multicolumn{1}{l|}{} & \multicolumn{3}{c|}{\textbf{Ranking 1}} & \multicolumn{3}{c|}{\textbf{Ranking 2}} & \multicolumn{3}{c|}{\textbf{Ranking 3}} \\ \hline
\multicolumn{1}{|c|}{\#} & \textbf{User} & \textbf{On-topic} & \textbf{Individual} & \textbf{User} & \textbf{On-topic} & \textbf{Individual} & \textbf{User} & \textbf{On-topic} & \textbf{Individual} \\ \hline
\multicolumn{1}{|c|}{1} & homesnutrition & X &  & johnneustadt & X & & johnneustadt & X &  \\ \hline
\multicolumn{1}{|c|}{2} & ficajones & X & X & jo\_millar27 & X & X & solutions777 & X & X \\ \hline
\multicolumn{1}{|c|}{3} & helenvweaver & X & X & hatchbrenner &  &  & kingste29344921 & X & X \\ \hline
\multicolumn{1}{|c|}{4} & spriggsnutri & X &  & nchawkes & X & X & daisylu1964 &  & X \\ \hline
\multicolumn{1}{|c|}{5} & critcarelthtr & X &  & moz0373runner & X & X & zakariamarsli & X & X \\ \hline
\multicolumn{1}{|c|}{6} & danielleroisin\_ & X & X & aimsonhealth & X & X & meowaaaaaa &  & X \\ \hline
\multicolumn{1}{|c|}{7} & mynameisandyj & X & X & wordsharkv5 &  & X & vecta67 &  & X \\ \hline
\multicolumn{1}{|c|}{8} & fionaliu92 & X & X & fullcircle\_play & X &  & cosfordfamily1 & X & X \\ \hline
\multicolumn{1}{|c|}{9} & ldpartnership & X &  & qsprivatehealth & X &  & hayleycorriganx &  & X \\ \hline
\multicolumn{1}{|c|}{10} & milaestevam1 &  & X & socialissp &  &  & jhbrasfie &  & X \\ \hline

\end{tabular}
	}
	\caption{Top-10 ranked users for ranking functions (\ref{eq:rank1}) and (\ref{eq:rank2}) and (\ref{eq:rank3}), with indication of whether the user is on-topic/off-topic and individual vs association/professional. Such categories are useful to evaluate the ranking functions.}
	\label{tab:rank1}
	\vspace{-20pt}
\end{table}

\begin{align}
\textit{Ranking 1:} ~ \mathit{R1}(u) & = \frac{1}{\sum_{u \in C} \mathit{IC}(u) + 1} \cdot \sum_{u \in C} \mathit{TF}(u) \label{eq:rank1} \\
\textit{Ranking 2:} ~ \mathit{R2}(u) & = \lvert \mathit{FR}(u) - 1 \rvert \cdot \left(\sum_{u \in C} \mathit{TA}(U) + \sum_{u \in C} \mathit{IC}(U)\right) \label{eq:rank2} \\
\textit{Ranking 3:} ~ \mathit{R3}(u) & = \lvert \mathit{FR}(u) - 1 \rvert \cdot \left(\sum_{u \in C} \mathit{TA}(U) + \frac{1}{\sum_{u \in C} \mathit{IC}(U) + 1}\right) \label{eq:rank3}
\end{align}
Function (\ref{eq:rank1}) is designed to promote users who are at the ``fringe'' of their community, while giving credit to generic on-topic activities during the contexts. 
To achieve this, \textit{Topical Focus} $\mathit{TF}$ is used as a positive contribution, while a large in-degree $\mathit{IC}$ reduces the score.
In contrast, function (\ref{eq:rank2}) penalises user popularity, i.e., by using the complement of \textit{Follower Rank} $\mathit{FR}$, while rewarding prominence inside communities (in-degree $\mathit{IC}$) and information spreading by also considering shared links (\textit{Topical Attachment} $\mathit{TA}$).
Function (\ref{eq:rank3}) combines ideas from both (\ref{eq:rank2}) and (\ref{eq:rank1}).

\begin{figure}[htb]
	\centering
	\begin{subfigure}{.49\textwidth}
		\centering
		\includegraphics[width=\textwidth]{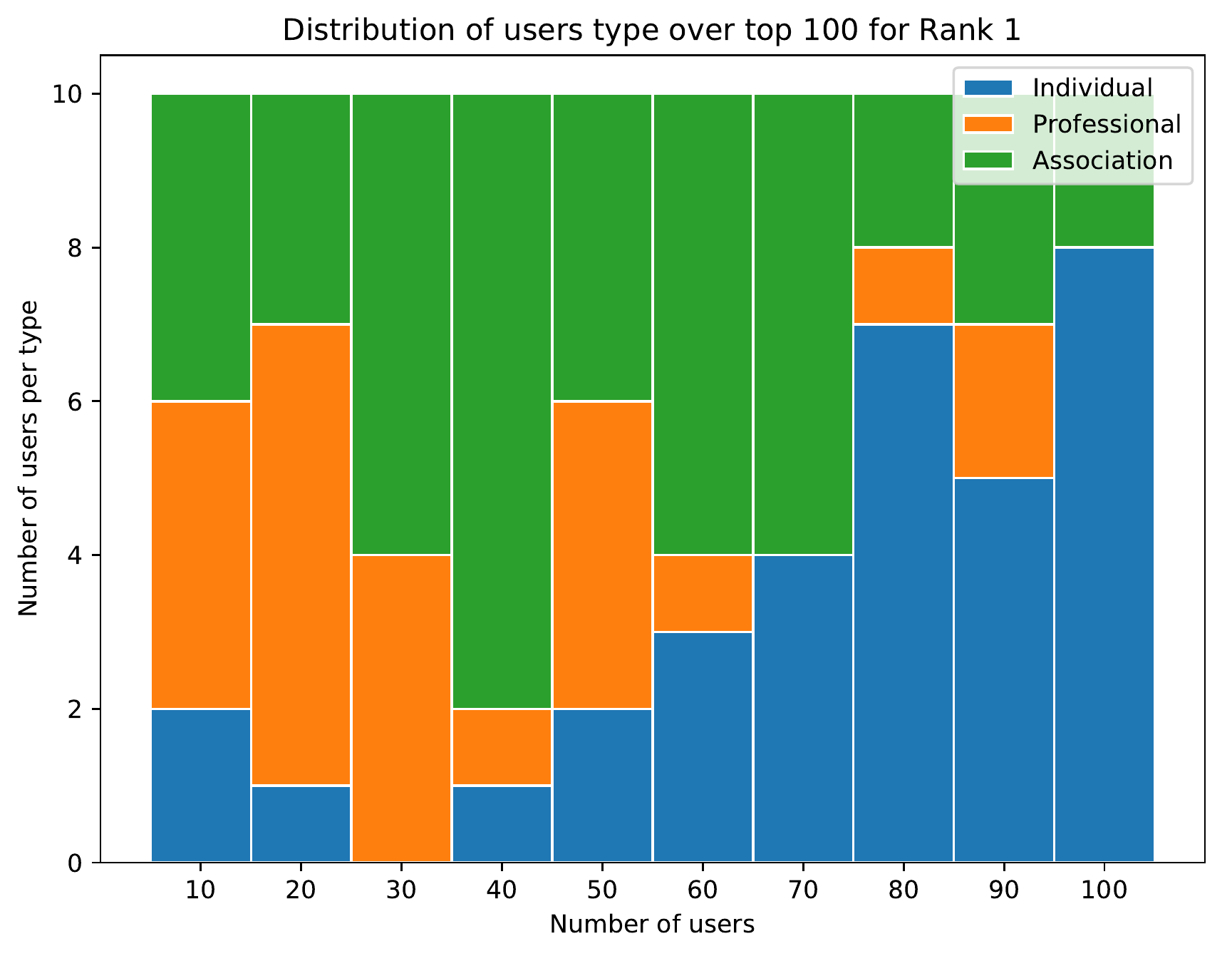}
		\caption{Rank 1: 33 are individuals, 23 are professionals, 44 are associations}
		\label{fig:rank1-distribution}
	\end{subfigure}
	\hfill%
	\begin{subfigure}{.49\textwidth}
		\centering
		\includegraphics[width=1\textwidth]{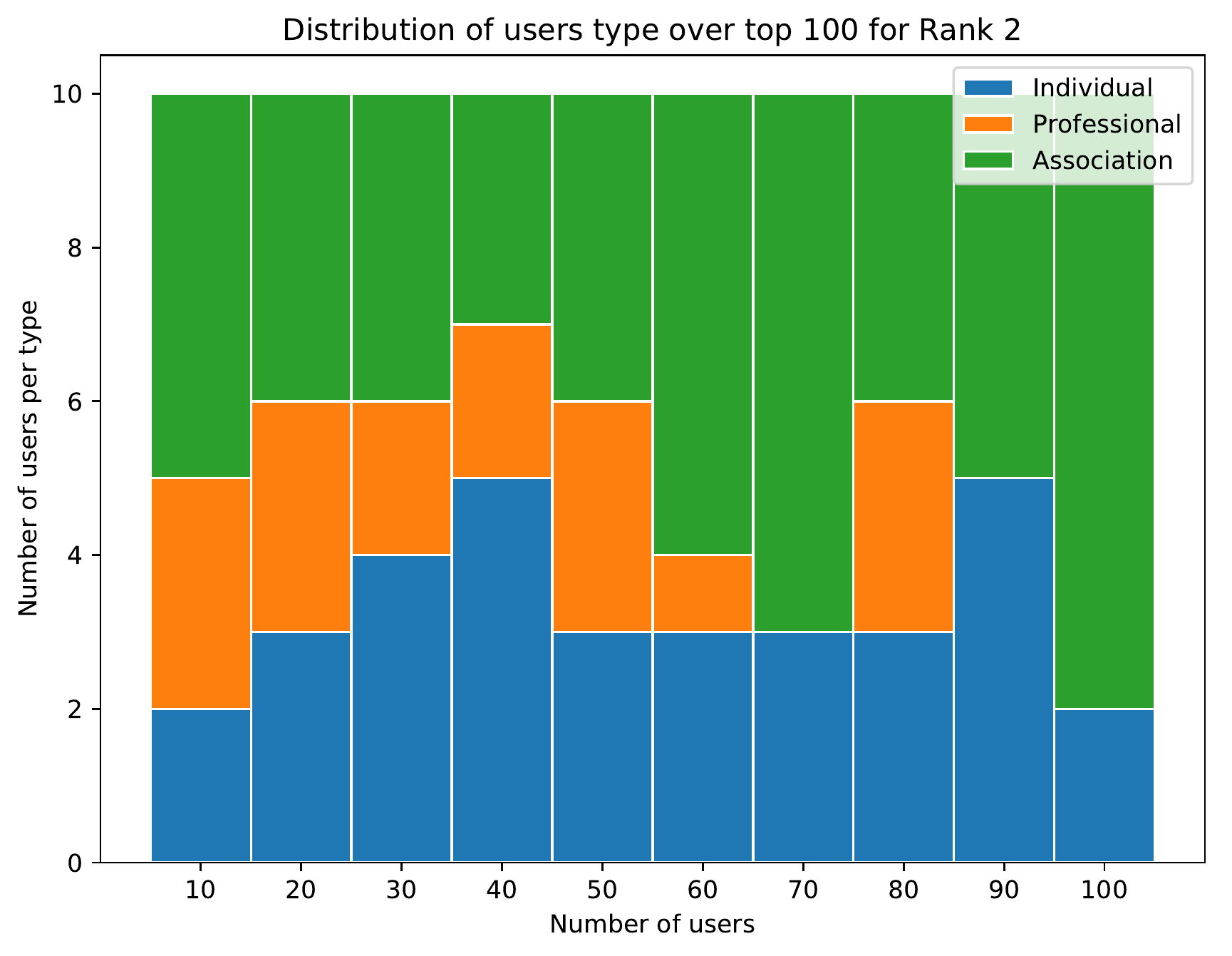}
		\caption{Rank 2: 33 are individuals, 17 are professionals, 50 are associations}
		\label{fig:rank2-distribution}
	\end{subfigure}
	\hfill%
	\begin{subfigure}{.49\textwidth}
		\centering
		\includegraphics[width=1\textwidth]{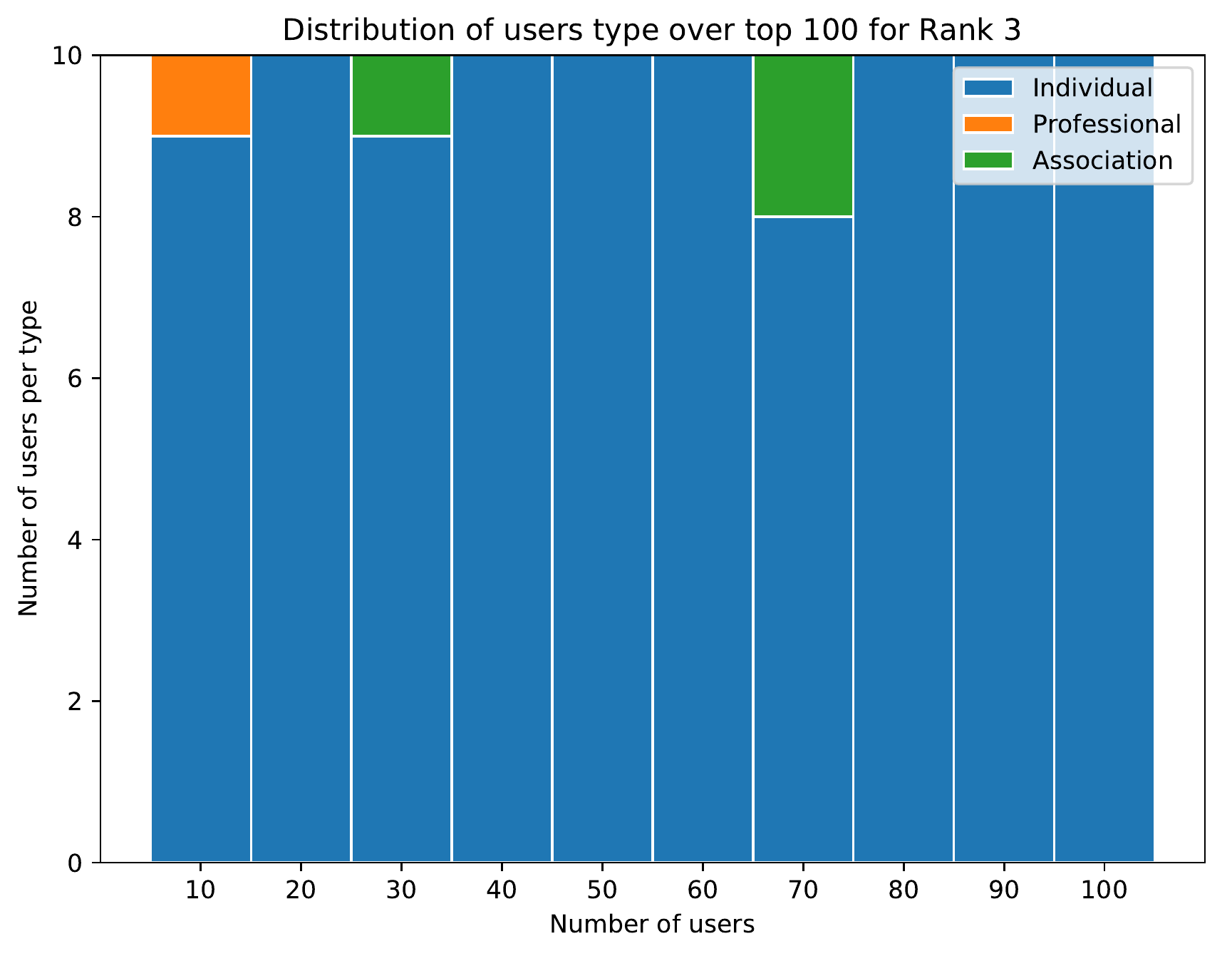}
		\caption{Rank 3: 96 are individuals, 1 are professionals, 3 are associations}
		\label{fig:rank3-distribution}
	\end{subfigure}
	\caption{Distribution of user types for top-100 users and for each ranking function.}
	\label{fig:ranks-distribution}
\end{figure}

The top-10 users for each ranking are reported in Tab.~\ref{tab:rank1}. To appreciate the effects of these functions, we have manually labelled the top-100 user profiles for each of the rankings, using a broad type classification as \textit{individuals} as opposed to \textit{institutional players} (associations, public bodies), or \textit{professionals}. 
The fractions of on-topic users are 86\%, 83\%, and 38\% for (\ref{eq:rank1}), (\ref{eq:rank2}), and (\ref{eq:rank3}) respectively.
Importantly, (\ref{eq:rank3}) identifies more individuals than institutions and professionals (96\%) than (\ref{eq:rank2}) and (\ref{eq:rank1}), both at 33\%p.
%
Also, repeat users are rewarded in both rankings. Users with $\mathit{FR}(u) = 0$ and $\mathit{min\_max(\lvert Tweets (u)\rvert) < 0.005}$ are considered not active and have been assigned lowest score.
Fig.~\ref{fig:ranks-distribution} shows the distribution of user types within the top-100 users for each of the three rankings, broken down into 10 users bins. We can see that individuals dominate in (\ref{eq:rank3}), and are fewer but emerge earlier in the ranks when (\ref{eq:rank2}) is used.
We plan to conduct user studies to establish useful analytics to be incorporated into our framework. 
    \vspace{-10pt}
\section{Conclusions and lessons learnt}
\vspace{-10pt}

Motivated by the need to find an operational definition of ``online activists'' that is grounded in well-established network and user-activity metrics, we have designed a Twitter content processing pipeline for progressively harvesting Twitter users based on their engagement with online socially-minded events, or campaigns, which we have called \textit{contexts}.
The pipeline yields a growing database of user profiles along with their associated metrics, which can then be analysed to experiment with user-defined user ranking criteria. The pipeline is  designed to select promising candidate profiles, but the approach is unsupervised, i.e., no manual classification of example users is provided.
We have empirically evaluated the pipeline on a case study, along with experimental scoring functions to show the viability of the approach. 

The design of the pipeline show that useful harvesting of interesting users can be accomplished within the limitations imposed by Twitter on its APIs.
The next challenge is to automate the discovery of new contexts so that the pipeline may continuously add new and update users in the database.
Only at this point will it be possible to   validate the entire approach, hopefully with help from third party users, on a variety of new context topics.


\vspace{-10pt}
	
    \begin{thebibliography}{10}

\scriptsize

\providecommand{\url}[1]{\texttt{#1}}
\providecommand{\urlprefix}{URL }
\providecommand{\doi}[1]{https://doi.org/#1}

\bibitem{Arnaboldi2013}
Arnaboldi, V., Conti, M., Passarella, A., Pezzoni, F.: {Ego networks in
  Twitter: An experimental analysis}. In: 2013 Proceedings IEEE INFOCOM. pp.
  3459--3464 (2013)

\bibitem{Biran2012}
Biran, O., Rosenthal, S., Andreas, J., McKeown, K., Rambow, O.: Detecting
  influencers in written online conversations. In: Proceedings of the Second
  Workshop on Language in Social Media. pp. 37--45. LSM '12, Association for
  Computational Linguistics, Stroudsburg, PA, USA (2012)

\bibitem{Bizid2018}
Bizid, I., Nayef, N., Boursier, P., Doucet, A.: {Detecting prominent microblog
  users over crisis events phases}. Information Systems  \textbf{78},  173--188
  (nov 2018)

\bibitem{Bizid:2015:PUD:2808797.2809411}
Bizid, I., Nayef, N., Boursier, P., Faiz, S., Morcos, J.: {Prominent Users
  Detection During Specific Events by Learning On- and Off-topic Features of
  User Activities}. In: Proceedings of the 2015 IEEE/ACM International
  Conference on Advances in Social Networks Analysis and Mining 2015. pp.
  500--503. ASONAM '15, ACM, New York, NY, USA (2015)

\bibitem{Bizid:2015}
Bizid, I., Nayef, N., Boursier, P., Faiz, S., Morcos, J.: {Prominent Users
  Detection During Specific Events by Learning On- and Off-topic Features of
  User Activities}. In: Proceedings of the 2015 IEEE/ACM International
  Conference on Advances in Social Networks Analysis and Mining 2015. pp.
  500--503. ASONAM '15, ACM, New York, NY, USA (2015)

\bibitem{doi:10.1080/14742830701497277}
Bobel, C.: ``i'm not an activist, though i've done a lot of it'': Doing
  activism, being activist and the ``perfect standard'' in a contemporary
  movement. Social Movement Studies  \textbf{6}(2),  147--159 (2007)

\bibitem{Bonacich2001}
Bonacich, P., Lloyd, P.: {Eigenvector-like measures of centrality for
  asymmetric relations}. Social Networks  \textbf{23}(3),  191--201 (jul 2001)

\bibitem{MATIC2011}
Booth, N., Matic, J.A.: Mapping and leveraging influencers in social media to
  shape corporate brand perceptions. corporate communications 16, 184-191.
  Corporate Communications: An International Journal  \textbf{16},  184--191
  (08 2011)

\bibitem{Cha2010MeasuringUI}
Cha, M., Haddadi, H., Benevenuto, F., Gummadi, K.P.: Measuring user influence
  in twitter: The million follower fallacy. In: ICWSM (2010)

\bibitem{Coscia:2012:DLD:2339530.2339630}
Coscia, M., Rossetti, G., Giannotti, F., Pedreschi, D.: Demon: A local-first
  discovery method for overlapping communities. In: Proceedings of the 18th ACM
  SIGKDD International Conference on Knowledge Discovery and Data Mining. pp.
  615--623. KDD '12, ACM, New York, NY, USA (2012)

\bibitem{Fisher2017}
Fisher, D.N., Silk, M.J., Franks, D.W.: The Perceived Assortativity of Social
  Networks: Methodological Problems and Solutions, pp. 1--19. Springer
  International Publishing, Cham (2017)

\bibitem{Kardara2015}
Kardara, M., Papadakis, G., Papaoikonomou, A., Tserpes, K., Varvarigou, T.:
  {Large-scale evaluation framework for local influence theories in Twitter}.
  Information Processing {\&} Management  \textbf{51}(1),  226--252 (2015)

\bibitem{IJoC1246}
Lotan, G., Graeff, E., Ananny, M., Gaffney, D., Pearce, I., Boyd, D.: The arab
  spring| the revolutions were tweeted: Information flows during the 2011
  tunisian and egyptian revolutions. International Journal of Communication
  \textbf{5}(0) (2011)

\bibitem{Missier2017}
Missier, P., McClean, C., Carlton, J., Cedrim, D., Silva, L., Garcia, A.,
  Plastino, A., Romanovsky, A.: {Recruiting from the Network: Discovering
  Twitter Users Who Can Help Combat Zika Epidemics}. In: Web Engineering: 17th
  International Conference, ICWE 2017, Rome, Italy, June 5-8, 2017,
  Proceedings. pp. 437--445. Springer International Publishing, Roma, Italy
  (2017)

\bibitem{7569535}
{Nargundkar}, A., {Rao}, Y.S.: Influencerank: A machine learning approach to
  measure influence of twitter users. In: 2016 International Conference on
  Recent Trends in Information Technology (ICRTIT). pp.~1--6 (April 2016)

\bibitem{Overbey2013}
Overbey, L.A., Greco, B., Paribello, C., Jackson, T.: {Structure and prominence
  in Twitter networks centered on contentious politics}. Social Network
  Analysis and Mining  \textbf{3}(4),  1351--1378 (dec 2013)

\bibitem{Pal2011}
Pal, A., Counts, S.: {Identifying topical authorities in microblogs}. In:
  Proceedings of the fourth ACM international conference on Web search and data
  mining - WSDM '11 (2011)

\bibitem{Poell2014}
Poell, T.: {Social media and the transformation of activist communication:
  exploring the social media ecology of the 2010 Toronto G20 protests}.
  Information, Communication {\&} Society  \textbf{17}(6),  716--731 (jul 2014)

\bibitem{6978961}
{Razis}, G., {Anagnostopoulos}, I.: Semantifying twitter: The influence tracker
  ontology. In: 2014 9th International Workshop on Semantic and Social Media
  Adaptation and Personalization. pp. 98--103 (Nov 2014)

\bibitem{RIQUELME2016949}
Riquelme, F., Gonzalez-Cantergiani, P.: Measuring user influence on twitter: A
  survey. Information Processing and Management  \textbf{52}(5),  949 -- 975
  (2016)

\bibitem{INFOMAP}
Rosvall, M., Bergstrom, C.T.: Maps of random walks on complex networks reveal
  community structure. Proceedings of the National Academy of Sciences of the
  United States of America  \textbf{105},  1118--23 (02 2008)

\bibitem{Schenk2011}
Schenk, C.B., Sicker, D.C.: {Finding Event-Specific Influencers in Dynamic
  Social Networks}. In: 2011 IEEE Third International Conference on Privacy,
  Security, Risk and Trust and 2011 IEEE Third International Conference on
  Social Computing. pp. 501--504 (2011)

\bibitem{Sousa2018}
Sousa, L., de~Mello, R., Cedrim, D., Garcia, A., Missier, P., Uchoa, A.,
  Oliveira, A., Romanovsky, A.: Vazadengue: An information system for
  preventing and combating mosquito-borne diseases with social networks.
  Information Systems  \textbf{75},  26 -- 42 (2018)

\bibitem{Youmans2012}
Youmans, W.L., York, J.C.: Social media and the activist toolkit: User
  agreements, corporate interests, and the information infrastructure of modern
  social movements. Journal of Communication  \textbf{62}(2),  315--329 (2012)

\bibitem{Zhao2011b}
Zhao, W.X., Jiang, J., Weng, J., He, J., Lim, E.P., Yan, H., Li, X.: {Comparing
  twitter and traditional media using topic models}. 33rd European Conference
  on IR Research, ECIR 2011 pp. 338--349 (2011)

\end{thebibliography}
%
	\vspace{-10pt}
    \section*{Acknowledgments}
    The authors would like to thank Prof. Carlo Piccardi at Politecnico di Milano, Italy, for his useful suggestions.

\end{document}